\documentclass{article}  
\usepackage{lajolla2006}
\usepackage{graphicx}
\frompage{000} \topage{000}                                              
\def\rightmark{Scaling Characteristics of Azimuthal Anisotropy at RHIC}
\def\leftmark{M.Issah and A. Taranenko for the PHENIX Collaboration}
 
\title{Scaling Characteristics of Azimuthal Anisotropy at RHIC} 
\authors{
{Michael Issah$^1$ and Arkadij Taranenko$^1$ , for the PHENIX Collaboration%
}\\[2.812mm]
{\normalsize
\hspace*{-8pt}$^1$ Department of Chemistry, State University of 
New York at Stony Brook, Stony Brook, New York 11794-3400, USA\\[0.2ex]  
}}

\abstract{ Recent differential measurements of elliptic flow are used to 
probe several hydrodynamic scaling predictions. Eccentricity scaling is 
observed for Cu+Cu and Au+Au collisions at $\sqrt{s_{NN}}=~200$ GeV, suggesting 
essentially complete thermalization of the high energy density matter produced 
in these collisions. An estimate of the speed of sound is also obtained from 
the eccentricity scaled elliptic flow. The predicted mass scaling is observed 
for particles with transverse kinetic energy KE$_T$ up to $\sim 1$ GeV. For 
KE$_T$ values above $\sim 1$ GeV, valence quark number scaling compatible with partonic 
degrees of freedom is observed.
}

 
\begin{document}
 
\maketitle
\setcounter{page}{1}

\section{Introduction}\label{intro}

Experimental measurements at the Relativistic Heavy Ion Collider (RHIC) have 
indicated the creation of matter having energy density in excess of what is 
required for the formation of a de-confined phase of quarks and gluons 
(QGP)\cite{Adler:2004zn,Adcox:2004mh,Fodor:2001pe}. The decay of this matter 
results in particle emissions which show large anisotropies in their azimuthal 
distributions relative to the reaction plane. Such anisotropies, which are 
commonly characterized by the second Fourier coefficient $v_2$, suggest that 
the created high-energy density matter rapidly thermalizes and its decay dynamics 
is driven by substantial pressure gradients throughout its dynamical evolution 
\cite{Shuryak:2004cy,Gyulassy:2004zy,Heinz:2005ja}. 

	This picture for the matter created in RHIC collisions is supported by 
the good agreement achieved between $v_2$ measurements 
\cite{Adcox:2002ms,Adler:2002pu,Adler:2003kt,Adams:2004bi,Adler:2004cj,Lacey:2005qq} 
 and those obtained form 
hydrodynamic model calculations \cite{Kolb:2000fh,Huovinen:2003fa,Csanad:2005gv}.
 However, the detailed properties of the matter 
(ie. its equation of state, transport coefficients, effective degrees of freedom, 
etc), are still lacking. Recently, several questions concerning the degree of thermalization 
of this matter have even been under debate \cite{Bhalerao:2005mm,Csanad:2005gv}.  
	
	In this contribution, we show that the scaling characteristics of azimuthal 
anisotropy provide an important framework for developing important insights relevant 
to the properties of the high energy density matter created in ultrarelativistic 
heavy ion collisions at RHIC. 

\section{Data \& data analysis}\label{analysis} 
The results presented here utilize data obtained by the PHENIX collaboration for 
the Au+Au and Cu+Cu colliding systems at $\sqrt{s_{NN}}$=200 GeV. A detailed
description of the PHENIX detector is given elsewhere~\cite{nim_1}. The procedures 
for tracking and event selection are described in Refs.~\cite{Adler:2003kt,Adler:2004cj}.
The azimuthal anisotropies relevant to this work were measured using the reaction 
plane method \cite{Danielewicz:1985hn,Poskanzer:1998yz,Ollitrault:1993ba}. This method 
measures the correlations between the azimuthal angle of the particles detected 
in the central arms of PHENIX and the azimuth of the reaction plane obtained with 
the two Beam-Beam Counters (BBC) located at $\mid{\eta}\mid \sim 3 - 3.9$ 
\cite{Adler:2003kt}. An important advantage of the large $\eta$ separation between 
the BBC's and the central arms of PHENIX is the expectation that the evaluation of the 
azimuth of the reaction plane is much less affected by possible non-flow 
contributions, especially those due to jets\cite{Adler:2003kt}.

The magnitude of these elliptic flow correlations was determined 
by evaluating the second Fourier coefficient $v_2$, of the azimuthal distribution 
of emitted particles relative to the azimuth of the reaction plane  
\cite{Danielewicz:1985hn,Poskanzer:1998yz,Ollitrault:1993ba}:
\begin{equation}
v_2\equiv\langle e^{2i(\phi-\Phi_{RP})}\rangle
=\langle \cos 2(\phi-\Phi_{RP})\rangle.
\label{rxn_plane}
\end{equation}
Here, $\phi$ is the azimuthal angle of a particle measured in the laboratory 
coordinate system and $\Phi_{RP}$ is an estimate of the azimuth of the reaction 
plane. A resolution factor was evaluated and used to correct for the dispersion of 
the measured reaction plane about the true reaction plane. 

Detailed measurements were performed over a broad range of centralities, $p_T$ and 
particle species. The results from these measurements are presented in the following
sections.

\section{Insights into the degree of thermalization}\label{thermalization} 
\subsection{Eccentricity scaling}\label{ecc_scaling} 

As mentioned earlier, the large $v_2$ values measured at RHIC 
\cite{Adcox:2002ms,Adler:2002pu,Adler:2003kt,Adams:2004bi,Adler:2004cj,Lacey:2005qq}  
are compatible with those obtained from hydrodynamic model 
calculations \cite{Kolb:2000fh,Huovinen:2003fa,Csanad:2005gv}. Local thermalization 
constitutes an important assumption for the legitimate use of hydrodynamics to describe 
the data. This being so, a crucial question is whether or not one can gain further 
insights into the degree of thermalization from the data. To do this, we recall here 
that an important prediction of the hydrodynamic model is 
that $v_2$ should (i) scale with eccentricity and (ii) be independent of the colliding 
system for a given eccentricity \cite{Bhalerao:2005mm,Csanad:2005gv,Ollitrault:1992bk}. 
The eccentricity $\epsilon = \langle y^2 \rangle -  
\langle x^2 \rangle / (\langle y^2 \rangle +  \langle x^2 \rangle)$, measures the asymmetry 
of the ``almond shaped" collision zone containing the high energy density matter created 
in the collision; $x$ and $y$ are, respectively, the short and long axes of the almond.

\begin{figure}[!htb]
\begin{center}
\includegraphics[width=0.55\linewidth]{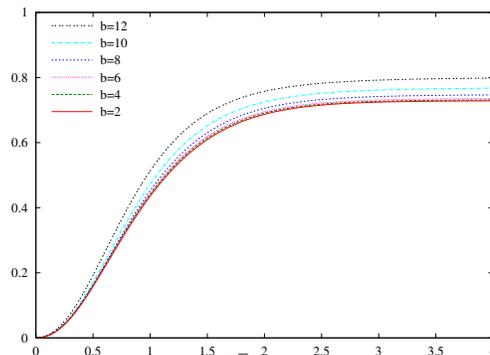}
\vspace*{-0.5cm}
\caption{$v_2/\epsilon$ as a function of $c_s (t-t_0)/\bar R$ for different
impact parameters. The plot is taken from \cite{Bhalerao:2005mm}.
\label{fig1}}
\end{center}
\end{figure}
%
	The eccentricity scaling prediction of hydrodynamics is illustrated 
in Fig.~\ref{fig1}.
The figure shows that, for a system of transverse size 
$\bar R$, elliptic flow develops over a time scale $\sim \bar R/\left\langle 
c_s\right\rangle$ for matter with a speed of sound $c_s$ \cite{Bhalerao:2005mm}. 
More importantly, it shows the variation of the scaled anisotropy $v_2/\epsilon$, 
as a function of the characteristic distance $c_s (t-t_0)/\bar R$, where 
$t_0$ is the thermalization time. The scaled results, which are shown for several  
impact parameters b or centralities, are taken from Ref.~\cite{Bhalerao:2005mm}. 
They indicate that $v_2/\epsilon$ is essentially independent of centrality except 
for the most peripheral collisions, where hydrodynamics is expected to break down.
This result provides a clear signature for hydrodynamic scaling which can be explored
with data. 

\begin{figure}[!htb]
\begin{tabular}{cc}
\includegraphics[width=0.5\linewidth]{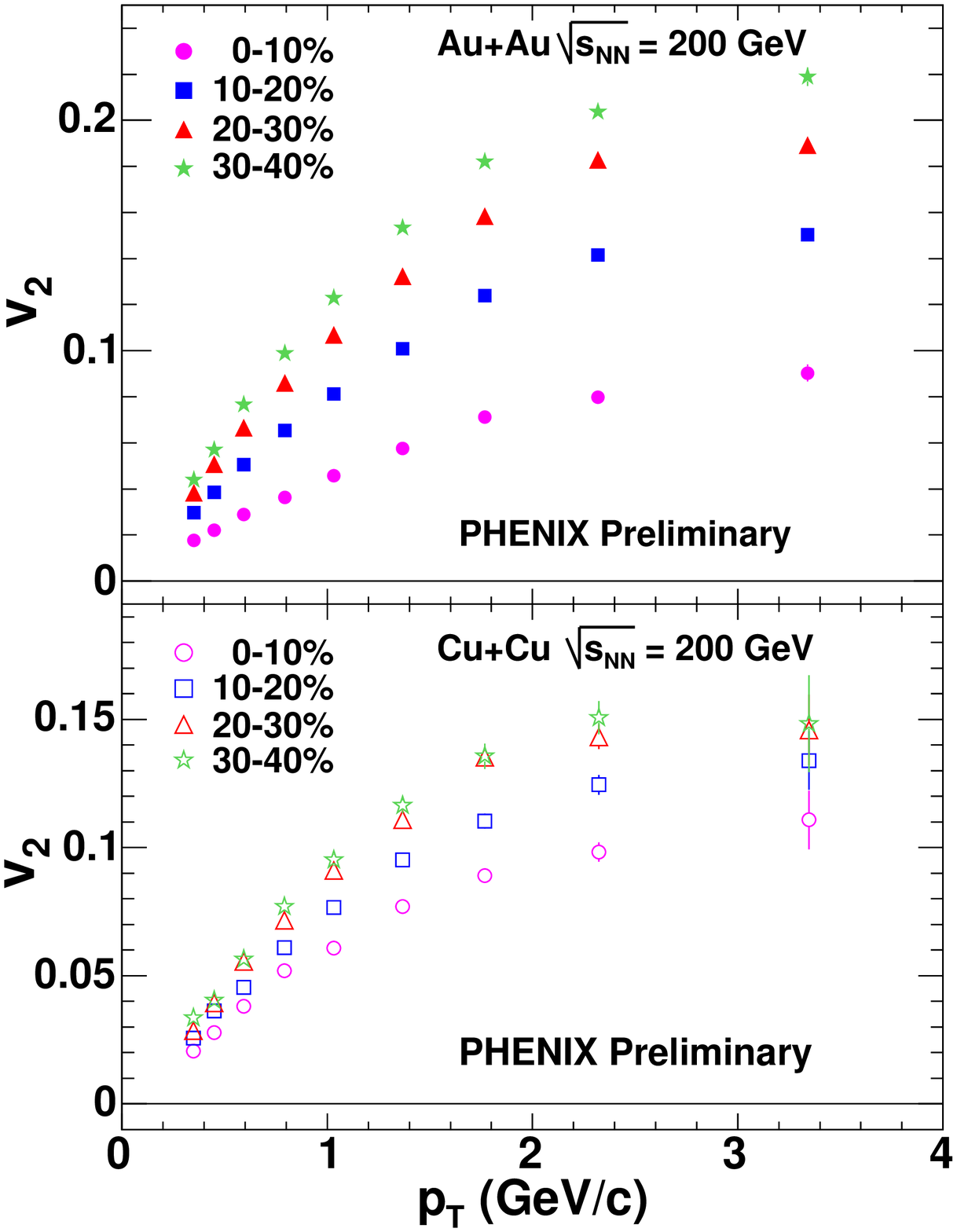} &
\includegraphics[width=0.5\linewidth]{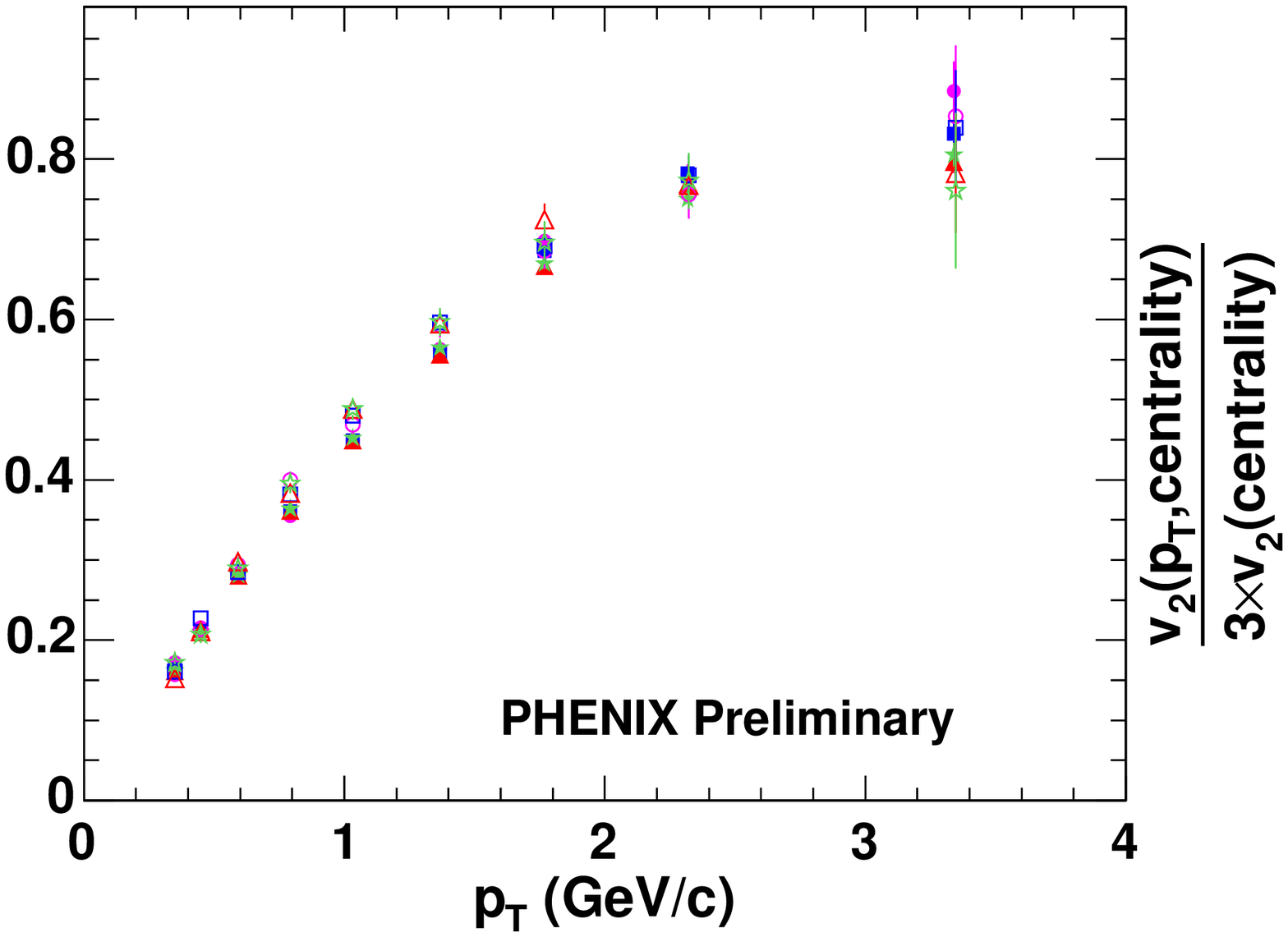}
\end{tabular}
\vspace{-0.5cm}
\caption{The left panels show $v_2$ vs. $p_T$ for several centrality selections 
for Au+Au (top) and Cu+Cu (bottom) collisions at $\sqrt{s_{NN}}=~200$~GeV.
The right panel shows $v_2(p_T)$ scaled by the $p_T$-integrated $v_2$ values for 
the same centralities indicated in the left panels.
\label{fig2}}
\end{figure}

	The left panels of Fig.~\ref{fig2} show recently measured differential 
elliptic flow $v_2(p_T)$ for several centralities in Au+Au and Cu+Cu collisions 
at $\sqrt{s_{NN}}$=200 GeV. For both systems, the characteristic increase of 
$v_2(p_T)$ as collisions become more peripheral is evident. In a hydrodynamic 
scenario, this increase is expected because $\epsilon$ grows as collisions become 
less central and larger values of $\epsilon$ lead to larger pressure gradients 
which drive the magnitude of collective flow. The crucial question here is 
whether or not these data show the eccentricity scaling predicted by hydrodynamic 
models.

	The right panel of Fig.~\ref{fig2} shows the scaled values of $v_2(p_T)$ 
for the same centralities shown in the left panel of the figure. To estimate the 
eccentricity, we use the fact that the $p_T$-integrated $v_2$ values vary linearly 
with centrality and are essentially proportional to $\epsilon$, i.e $\epsilon 
\sim k \times v_2$ over a broad range of centrality selections. Here, it should 
be emphasized that this proportionality has been experimentally verified in Au+Au 
collisions \cite{Adcox:2002ms,Adler:2002pu}. From Glauber model estimates \cite{Adcox:2002ms} 
of $\epsilon$, we find that $k \sim 3$ for the cuts employed in this analysis. 
This factor is reflected in the right panel of Fig.~\ref{fig2} where we have used 
$v_2(p_T)$/3$v_2$(centrality) for the scaled anisotropy. This ratio has the added 
advantage of being less affected by the systematic errors associated with our 
$v_2$ measurements since they cancel out.

The scaled $v_2$ values shown in the right panel of Fig.~\ref{fig2} indicate rather 
good scaling for the full range of centralities (or $\epsilon$) presented. This is 
clearly consistent with the eccentricity scaling predictions of hydrodynamics. 
Equally important is the fact that these results also confirm the expected independence 
on colliding system size. 

We interpret these findings as an indication that essentially full thermalization 
is achieved over a broad set of these Au+Au and Cu+Cu collisions \cite{Lacey:2005qq,Bhalerao:2005mm}.
The observation of the independence of $v_2/\epsilon$ on the system size also 
satisfies the requirement of scale invariance which is a basic feature of perfect 
fluid hydrodynamics. 

	Given these observations, it is important to ask whether or not one can use 
them to extract constraints for the properties of the high energy density matter.
This is pursued in the next section.

\subsection{An estimate of the speed of sound from scaled $v_2$ data}\label{speedofsound} 

	An important input for the hydrodynamic model is the equation of state (EOS).
The EOS relates the pressure $p$ to the energy density $\varepsilon$. For a simple 
EOS, $p = {c_{s}}^2 \varepsilon$. Since $v_2$ is driven by pressure gradients in the 
hydrodynamical model, the value of the speed of sound $c_s$ can greatly influence the 
magnitude of $v_2$.

\begin{figure}[!htb]
\begin{tabular}{cc}

\hspace*{-0.4cm}\includegraphics[width=0.5\linewidth]{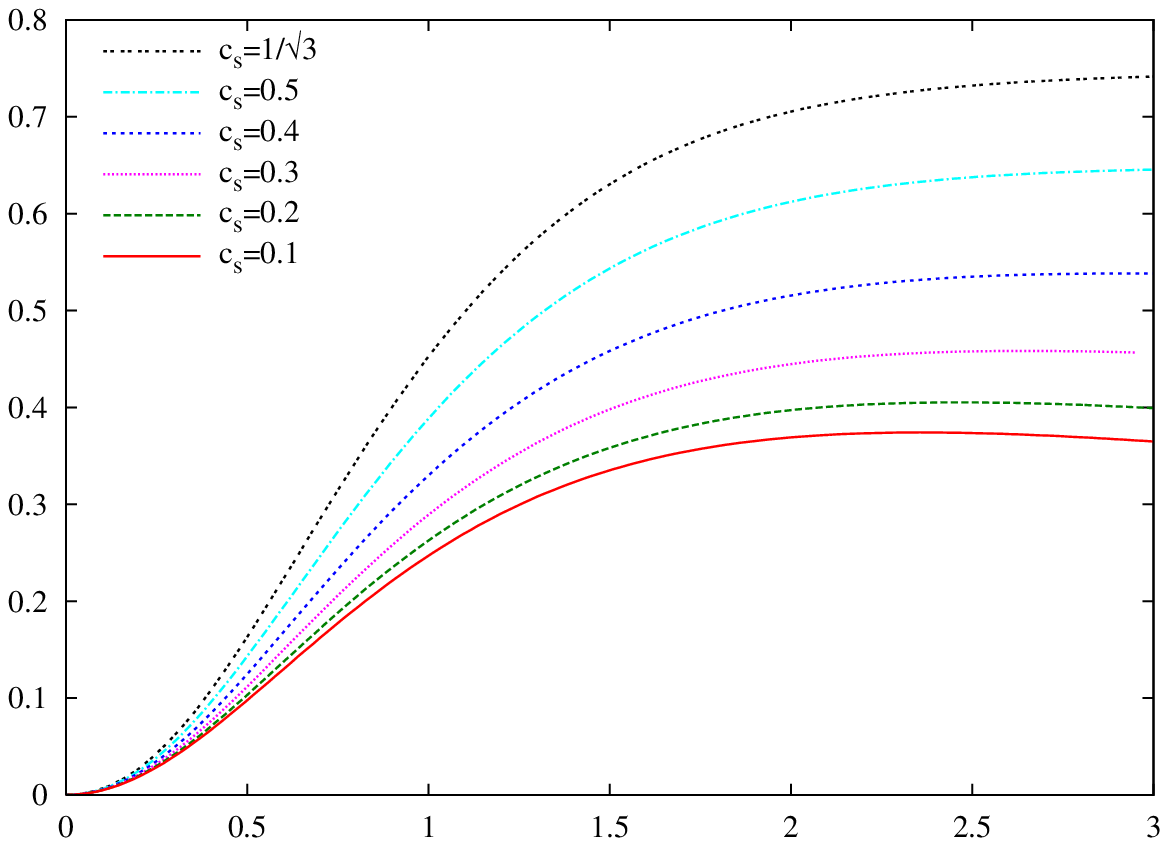} &
\hspace*{-0.5cm}\includegraphics[width=0.55\linewidth]{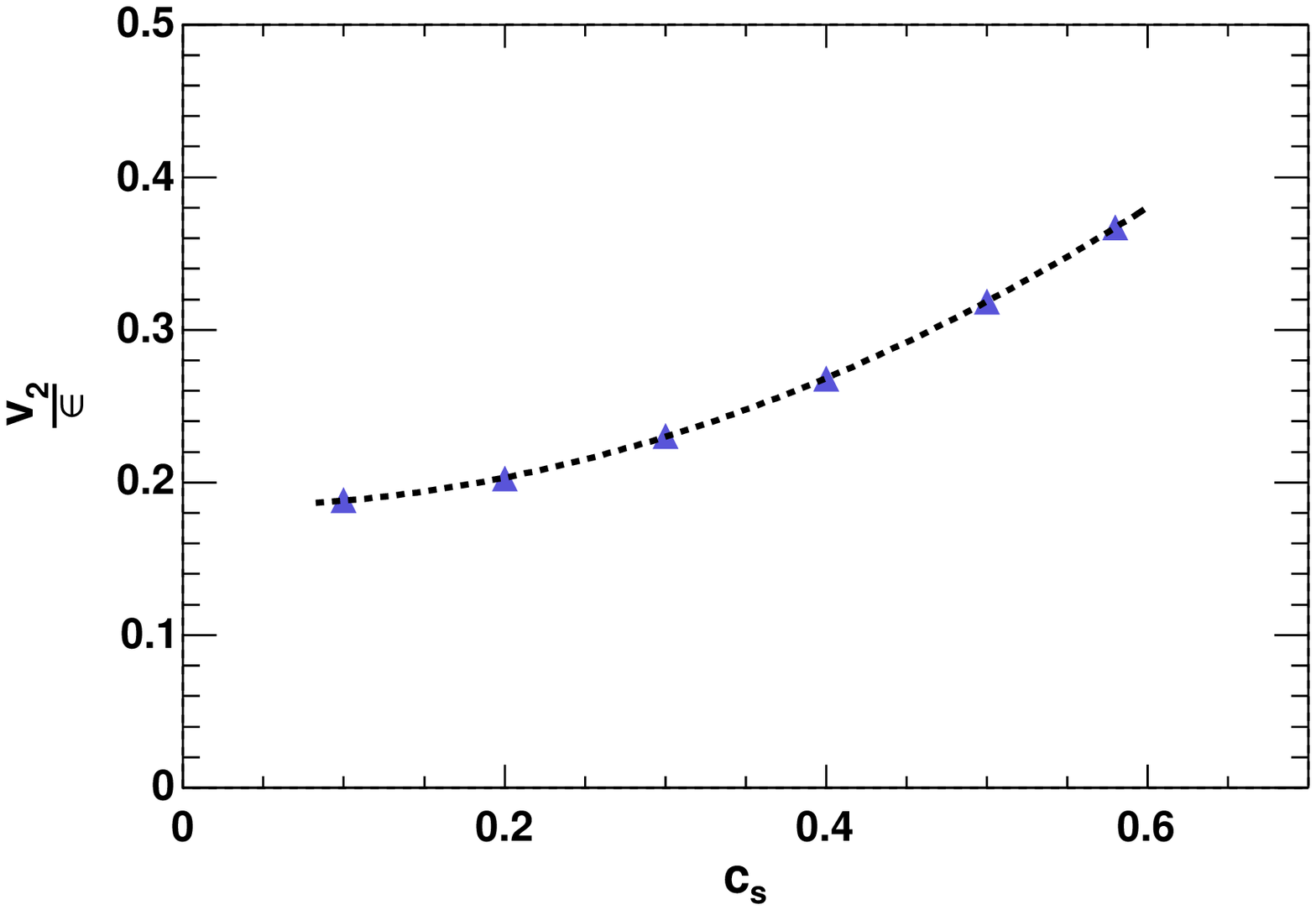}

\end{tabular}

\vspace*{-0.5cm}
\caption{The left panel shows $v_2/\epsilon$ as a function of $c_s (t-t_0)/\bar R$ for 
various values of $c_s$. The calculation was performed for Au+Au collisions with impact 
parameter $b=8$~fm at $\sqrt{s_{NN}}=200$~GeV \cite{Bhalerao:2005mm}. The right panel 
gives a summary of the dependence of $v_2/\epsilon$ on $c_s$. The dotted curve is 
represent a fit to the simulated data.}
\label{fig3}
\end{figure}

The results from a recent study of the influence of $c_s$ on the magnitude of 
$v_2/\epsilon$ \cite{Bhalerao:2005mm,Ollitrault:1992bk} is shown in Fig.~\ref{fig3}.
The figure indicates that the magnitude of $v_2/\epsilon$ is rather sensitive 
to the assumed value for $c_s$. This sensitivity is made more transparent in 
the right panel of Fig.~\ref{fig3} where we have summarized the results.

	An estimate of the speed of sound can be made by comparing the 
eccentricity scaled $v_2$ values shown in Fig.~\ref{fig2}, with the model predictions 
shown in Fig.~\ref{fig3}. It is noteworthy that we have checked that the 
eccentricities used in the model are comparable to Glauber based model 
estimates \cite{Adcox:2002ms} used in our evaluation of $k$. It should be pointed 
out that the $v_2/\epsilon$ values in the left panel of Fig.~\ref{fig3} 
differ from those in the right panel by a factor of 2. This factor was applied 
to take account of a well known difference in the procedure used to evaluate 
$v_2$ from the data and theory \cite{Bhalerao:2005mm}. A comparison between the 
right panels of Figs.~\ref{fig2} and \ref{fig3} gives the estimate $c_s$ $\sim$ 
0.35 $\pm$ 0.05 for a $\langle p_T \rangle$ $\sim$ 0.45 GeV/c \cite{Adler:2003cb}. 
This value of $c_s$ is indicative of a soft equation of state \cite{Adler:2004cj}), 
with the matter possibly spending little time, during its evolution, in a region where 
$c_s$ = 0.

\subsection{Transverse kinetic energy scaling}\label{transke} 

Given the observed eccentricity scaling, one can try to explore other
hydrodynamic scaling scenarios. For instance, the variation of $v_2$ with 
transverse kinetic energy KE$_T$ = $m_{T}-m$, where $m_T=\sqrt{p_T^2+m^2}$ is the transverse 
mass of the particle, is a good candidate for study. The use of this variable is intuitive 
because pressure gradients, which give rise to azimuthal anisotropy, lead to 
collective transverse kinetic energy of the emitted particles. 

\begin{figure}[!htb]
\begin{center}
\includegraphics[width=0.85\linewidth]{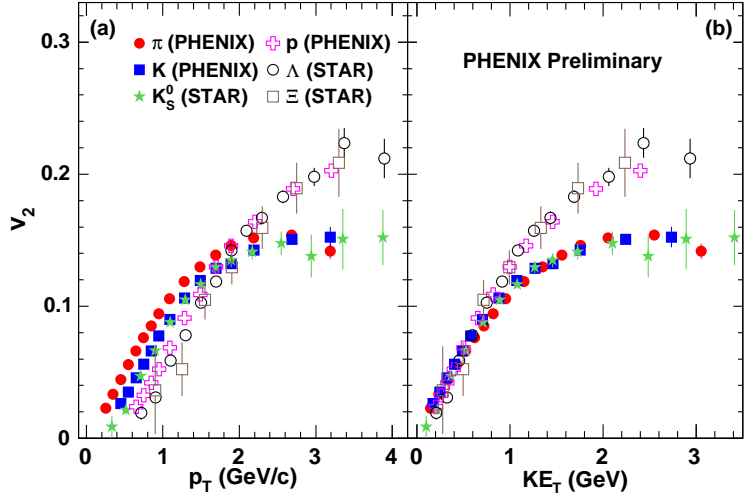}
\vspace*{-0.5cm}
\caption{Variation of $v_2$ with (a) $p_T$ and (b) KE$_T$ for 
selected hadrons for minimum bias collisions. The pion ($\pi^{\pm}$), kaon 
($K^{\pm}$) and (anti-)proton (($\bar{p}$)p) results are from PHENIX 
\cite{Masui:2005aa} and the neutral kaon $K_{0}^{S}$, lambda $\Lambda$ 
and cascade $\Xi$ results are from STAR \cite{Adams:2005zg}.\label{fig4}}
\end{center}
\end{figure}

Figure~\ref{fig4} shows the dependence of $v_2$ on $p_T$ and KE$_T$ (respectively) 
for several identified particles. The values of $v_2$ for charged pions ($\pi^{\pm}$), 
kaons ($K^{\pm}$) and (anti-)protons (($\bar{p}$)p) are preliminary data 
from the PHENIX Collaboration \cite{Lacey:2005qq,Masui:2005aa} whereas those for $K_{0}^{S}$, 
$\Lambda$ and $\Xi$ are published data from STAR Collaboration \cite{Adams:2005zg}. 
The STAR $v_2$ values were multiplied  by a factor of 1.1 to account for a small 
difference in the mean centrality for minimum bias 
events measured by the two experiments. In the left panel of Fig.~\ref{fig4}, one 
can observe mass ordering of $v_2(p_T)$ for $p_T$ up to $\sim 2$ GeV/c. That is, for 
a given $p_T$, the larger the mass of a particle, the smaller its $v_2$. This pattern 
has been identified as a signature for hydrodynamic expansion \cite{Huovinen:2001cy}. 

In the right panel of Fig.~\ref{fig4}, $v_2$ is plotted as a function of $KE_{T}$. 
Here, it can be seen that $v_2(KE_T)$ shows the expected scaling with KE$_{T}$ up to 
$\sim 1$~GeV. For larger values of  KE$_T$, the scaling breaks. However, a rather 
striking trend emerges : the mesons and baryons scale independently for the whole 
KE$_{T}$ range. This pattern suggests that the constituents of which baryons and 
mesons are made also flow. We pursue this lead via an additional scaling test in 
the next section.

\section{What are the degrees of freedom in the high energy density matter ?}\label{degfree} 

The observation in Fig.~\ref{fig4} that mesons and baryons scale in two separate 
branches for KE$_T$ values above $\sim 1$ GeV suggests that the constituents of 
the hadrons flow. If this is indeed the case, then the data should reflect 
constituent quark number scaling. Indeed, several models based on the coalescence or 
recombination of constituent quarks have been developed \cite{Vo02,Fr03a,Gr03a,MV03,Hw03a}. 
In these models, mesons and baryons obtain their elliptic flow from the elliptic flow 
of partons in an additive way, so that we have the simple relations: $v_{2,M}(p_{T}) \simeq$
2 $v_{2,p}(p_{T}/2)$ and $v_{2,B}(p_{T}) \simeq$ 3 $v_{2,p}(p_{T}/3)$, where $v_{2,M}$, 
$v_{2,B}$ and $v_{2,p}$ are the $v_2$ of a meson, baryon and parton respectively. The 
underlying assumption of this picture is that each parton carries the same fraction of 
the hadron elliptic flow. 

\begin{figure}[!htb]
\begin{center}
\includegraphics[width=0.85\linewidth]{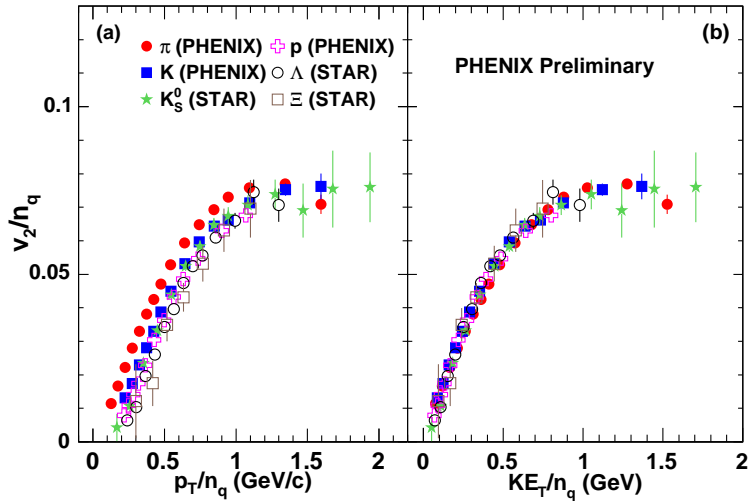}
\vspace*{-0.5cm}
\caption{Variation of $v_2$/$n_{q}$ with $p_T$/$n_{q}$ (left panel) 
and with KE$_{T}$/$n_q$(right panel) for different hadron species for minimum bias 
collisions.
\label{fig5}}
\end{center}
\end{figure}

The results from constituent quark number scaling are shown in Fig.~\ref{fig5}. Here the 
values for $v_2$, $p_T$ and KE$_{T}$ for different particle species are divided  by 
their respective number of quarks for mesons ($n_q$=2) and baryons ($n_q$=3). The 
left panel shows the commonly exploited quark number scaling ansatz: 
$v_2$($p_T$)/$n_{q}$ vs $p_T$/$n_{q}$. 
The left panel shows poor scaling for $p_{T}/n_{q}$ $\leq$ 1 GeV/c and better scaling, 
within errors, for $p_T$/$n_q$ $\geq$ 1.3 GeV/c. In contrast, the right panel shows 
that $v_2$(KE$_T$)/$n_{q}$ scales over the whole range of KE$_{T}$/$n_q$. This observation 
can be interpreted as a strong indication that the pertinent degrees of freedom in the 
flowing high energy density matter are linked to the number of constituent quarks of 
the hadron. 

\begin{figure}[!htb]
\begin{center}
\includegraphics[width=0.85\linewidth]{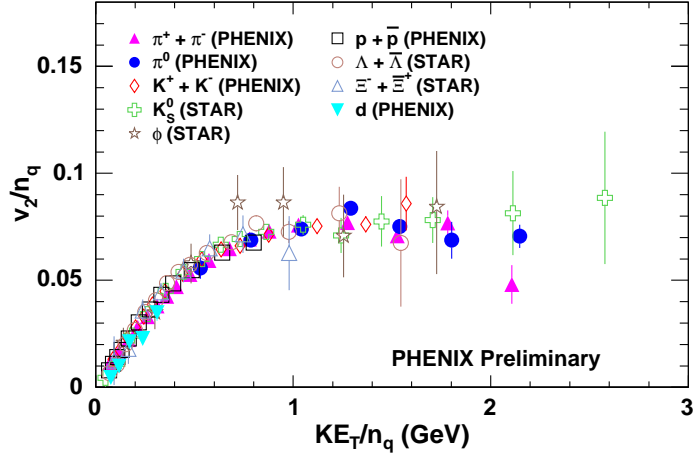}
\vspace*{-0.5cm}
\caption{Variation of $v_2$/$n_{q}$ with KE$_{T}$/$n_q$ for a 
comprehensive set of hadrons.\label{fig6}
}
\end{center}
\end{figure}

Fig.~\ref{fig6} shows the scaling with KE$_{T}$/$n_q$ for a more comprehensive set of 
hadron species. The $\phi$ meson and deuteron $v_2$ values are taken from \cite{Oldenburg:2005er}
and \cite{Lacey:2005qq} respectively. It appears that the scaling holds even for a heavy meson 
like the $\phi$ and for the deuteron ($n_q$=6), lending support to the idea that 
the different particle species emerge from a common flow field.

\subsection{Quark number scaling for different centrality selections}\label{centdep} 
 
\begin{figure}[!htb]
\begin{center}
\includegraphics[width=0.85\linewidth]{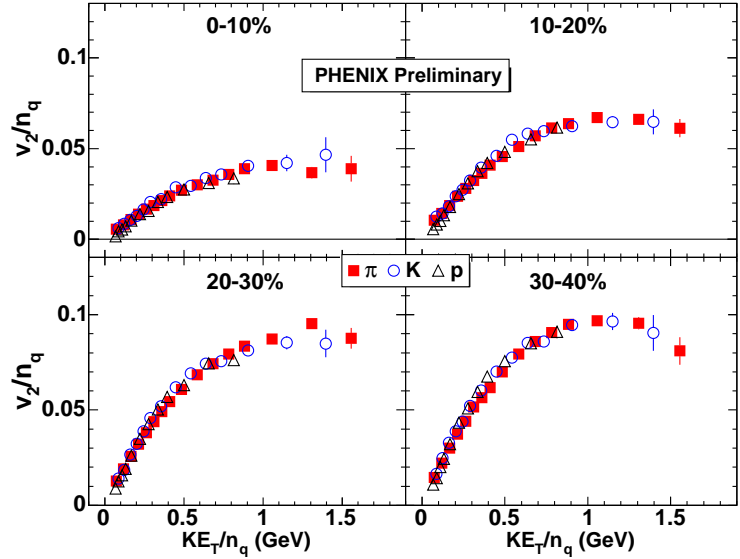}
\vspace*{-0.5cm}
\caption{Variation of $v_2$/$n_{q}$ with KE$_{T}$/$n_q$ for $\pi^{\pm}$, 
$K^{\pm}$ and p($\bar{p}$) for selected centralities for Au+Au collisions 
at $\sqrt{s_{NN}}=~200$ GeV.\label{fig7}} 
\end{center}
\end{figure}

Hydrodynamics is expected to break down for the most peripheral collisions and for 
relatively high $p_{T}$\cite{Kolb:2000fh}. Consequently, it is important to  
investigate the evolution of the scaling picture as the centrality of 
collisions is varied and the $p_T$ of the particles is increased. Such an analysis 
is still under intense investigation. Nonetheless, we show preliminary scaling results 
for a select set of centralities in Fig.~\ref{fig7}. Interestingly, quark number scaling 
of $v_2$ and KE$_{T}$ holds rather well for the range of centralities presented. It 
will be interesting to see if this scaling breaks for more peripheral collisions and 
at high $p_T$.

\section{Conclusions}\label{conclusion}

In summary, we have presented results for the scaling properties of azimuthal 
anisotropy at RHIC. The eccentricity scaled $v_2$ values for Cu+Cu and Au+Au at
$\sqrt{s_{NN}}=~200~GeV$ are similar as a function of $p_T$, showing that scale 
invariance of azimuthal anisotropy is satisfied. This suggests essentially complete 
thermalization of the high energy density matter produced in these collisions. 
The variation of $v_2$ with transverse relativistic kinetic energy KE$_{T}$ shows 
the predicted hydrodynamic mass scaling at low KE$_{T}$. However, it evolves into 
a differentiation of the mesonic and baryonic flow character of particles produced 
with high KE$_{T}$. This suggests that KE$_{T}$ is a good variable to 
study the dynamics of the produced matter. Further scaling of $v_2$ and KE$_{T}$ by 
the number of valence quarks leads to a unified scaling behavior over the whole range of 
KE$_{T}$/$n_q$ spanned by the data. The observations from eccentricity and 
transverse kinetic energy scaling point to the production of thermalized high energy 
density matter with partonic degrees of freedom.
 
\section{Acknowledgements}\label{ack}

We thank Hiroshi Masui and Roy Lacey for their invaluable contributions and helpful discussions.

\vfill\eject
\end{document}